\documentclass[rmp,amssymb,showpacs,showkeys,12pt,preprint]{revtex4}
\usepackage[T1]{fontenc}
\usepackage{graphicx}
\RequirePackage{times}
\RequirePackage{mathptm}
\bibpunct{[}{]}{,}{a}{,}{,}%
\begin{document}

\title{Aesthetic complexity\footnote{
Presented at the
{\em Time's Up Laboratories Data Ecology Workshop II}, Linz, Austria, May 2006,
as well as at the
{\em Alpbach Technology Forum 2007 --- Emergence in Science and Technology --- The Five Senses}, Alpbach, Tyrol, Austria, August 2007
}
}

\author{Karl Svozil}
\email{svozil@tuwien.ac.at}
\homepage{http://tph.tuwien.ac.at/~svozil}
\affiliation{Institute for Theoretical Physics, Vienna University of Technology, \\
Wiedner Hauptstra\ss e 8-10/136, A-1040 Vienna, Austria}

\begin{abstract}
Aesthetics, among other criteria, can be statistically examined in terms of the complexity required for creating and decrypting a work of art.
We propose three laws of aesthetic complexity.
According to the first law of aesthetic complexity, too condensed encoding makes a decryption of a work of art impossible and is perceived as chaotic by the untrained mind, whereas too regular structures are perceived as monotonous, too orderly and not very stimulating.
Thus a necessary condition for an artistic form or design to appear appealing is its complexity to lie within a bracket between monotony and chaos.
According to the second law of aesthetic complexity, due to human predisposition, this bracket is invariably based on natural forms; with rather limited plasticity.
The third law of aesthetic complexity states that aesthetic complexity trends are dominated by the available resources, and thus also by cost and scarcity.
\end{abstract}

\pacs{89.20.-a,89.75.-k,01.70.+w}
\keywords{Interdisciplinary applications of physics, complex systems, philosophy of science}
\maketitle

\begin{quote}
\begin{flushright}
{\footnotesize
dedicated to Hans Frank d. J\"ungeren\\
painter, teacher and friend}
\end{flushright}
\end{quote}

\section{Communicating art through encryption and decryption}

Every artist, in order to be able to create and communicate an artistic idea, has to express that idea in a way of form and design, so that the audience can interprete this form and design and get the artistic idea expressed in it.
From this point of view, art is perceived as a process of encryption and decryption of an artistic idea.
This is a Platonist, Kantian concept, because it is assumed that artistic ideas can never be communicated directly, but rather have to be encoded into some kind of symbolism.

Stated differently, no work of art exists without encryption and decryption.
In particular, every artistically encrypted pattern and form will eventually have to be decrypted by an audience.
Conversely, no form or design exists without an idea.
Thus, art occurs ``in the mind'' of its creators as well as of its recipients.
Both groups are ``artits of sorts,'' the difference being the difficult part of the creation as compared to the relative ease of its consumption.

A symbol can be anything communicable through the mind (brain?) --- (external?) world interface; i.e., through our senses.
Examples of artistic symbols are paints, colors, textures, tones, multi dimensional dimensional forms, touches, pixels on a digital screen, smells, as well as melodies and ornaments.
Every artistic symbolism depends, among other things,
on the idea expressed, as well as
on the preferred artistic style, which in turn is subject to tradition and historic developments, and also on the receptive human organs, including the brain functions interpreting nerve spiking activities from the sensory organs.

The idea expressed may be any kind of thought or emotion.
It may, for instance, consist of the feelings encountered in the first days of early spring,
the ``love that flows on a summer day,''
the wonder of our existence (rather than non-existence),
feelings of estrangement versus security,
and the golden sunsets on the sea or on an alpine mountain top at summer ends.
It may also be the vision of a tiny little program humming and jerking out patterns.

Indeed, the design itself may become the subject of an artistic idea.
So, designs and artistic ideas may merge and thus may no longer be perceived separately, or as vertical layers of the ``protocol'' of art.

\section{Dependence on complexity}

In what follows, we attempt to derive necessary, yet not entirely sufficient, statistical criteria for the aesthetic perception of art and its formal design expressions.
Particular consideration will be given to the amount of effort necessary to encrypt and decrypt the artistic form and design.

The  \textit{first law of aesthetic complexity} can be stated as follows:
\begin{quote}
{\em
The aesthetics of artistic forms and designs depend on their complexity.
Too condensed coding makes a decryption of a work of art impossible and is perceived as chaotic by the untrained mind, whereas too regular structures are perceived as monotonous, too orderly and not very stimulating.
}
\end{quote}

According to the first law of aesthetic complexity, the more complex a pattern in terms of description and production, the more difficult is its decryption.
Because if the decryption comes too fast and easy the result will be boredom;
conversely, if the decryption is too difficult, the result will be perplexity and irritation.

The first law  of aesthetic complexity has been introduced heuristically.
Yet its justification should ultimately come from a consideration of the human perception; in particular from its neuronal foundations.

There are several types of complexities, some measuring the length of the shortest algorithm or description of an artistic design, some measuring the amount of time and space required by its decryption.
We shall shortly discuss them now.

\textit{Descriptive complexity} can be characterized by the algorithmic information content
\cite{chaitin2,chaitin3,calude:02};
i.e., by the length of the shortest program capable to generate that pattern or form.
At one extreme, plain structures easily producible algorithmically appear monotonous.
At the other extreme, totally stochastic structures
appear irritating. That is to say, where patterns are simple and easily
recognized, the person experiencing them quickly loses interest; and equally
true, where there is no recognizable pattern at all, the person will experience frustration and lose interest in the apparent randomness.

\textit{Computational complexity}~\cite{calude1,bennett-utm}
is a measure for the amount of time and
memory (space) required to generate the pattern or form from the algorithm.
For example, a very short subroutine of only a few lines can generate a very
large pattern or form, but it may take a very large amount of time and
memory to accomplish this. The resulting pattern, then, is descriptionally
simple, but computationally complex.
Indeed, more formally, the time required for a decryption of a behavioral pattern created by a program of length $n$ may grow faster than any computable function of $n$;
it is only bounded~\cite{rado,chaitin-ACM,dewdney,brady} by the Busy Beaver function $\Sigma (n+O(1))$ of the sum of $n$ and the order of $1$.

\textit{Logical depth} \cite{ch6,bennett1,bennett-utm}
is a variant of computational complexity.
Heuristically speaking, it is a measure of the resources, in particular of the execution time, required to encode a design by its {\em canonical}, i.e., shortest length, program.

Art takes place in a ``bracket'' or region between monotony and irritation, between order and chaos.
Of course, the mere absence of monotony and randomness in the design is no sufficient criterion for art, but it can be safely stated that it is a necessary one.
Any attempt to push the artistic boundaries either towards monotony or towards
stochasticity must consider the human mind, which might not be sufficiently adaptive to cope with the results.

Consider the extremes of white noise and brown noise. White
noise is a type of noise that is produced by stochastically combining with equal weight all different frequencies together.
No correlation, no time dependence exists between its pieces.
White noise is thus characterized by a constant frequency spectrum $1/f^0$ and is too stochastic and random to be perceived as music; it is extremely irritating to most human ears.

In an apparent lack of harmony~\cite{Cage-autobio}, some composers and painters have intentionally introduced a great deal of more or less uncorrelated indeterminacy and noise~\cite{Sangild}.
By that method, even dilettantes or automata ``excel'' in the arts.
This style, which has its precursors in the futuristic~\cite{Russolo-1913}, dadaistic and surrealistic~\cite{breton:30,breton:33} (without reference to the subconscious) movements,
leaves many people irritated and perplexed.

With regards to randomness and determinacy, there appears to be a paradox: the more information a design and an artistic form contains, the more random it appears.
A very high information density requires very high resources for its decryption.
In the extreme, the most compact encoding results in pure randomness.
One provable example of this fact is the halting probability $\Omega$~\cite{chaitin2,chaitin3,calude:02}.

At the other bound of the artistic complexity bracket, we have brown noise---which takes its name from Brownian motion;
with a frequency spectrum $1/f^2$.
In this ``random walk'' type behavior, each event is based on the preceding events.
The resulting high correlations make Brown noise appear monotonous and boring.

At the mid-point between these extremes of noise, we find what we might term
``communicable music,'' which can be statistically characterized by a frequency spectrum of roughly about $1/f$.
This type of ``noise'' may also be termed ``fractal'' or self-similar ``noise''~\cite{voss-75,voss-78,gard-78,taylor-99}.

The term ``noise,'' of course, was coined to describe sound, but the
statistical analysis is easily applied to any mode of perception involving other forms of artistic designs and
pattern recognition~\cite{bovill}. For examples of ``noise'' in graphic or visual art, we
may look to modernist paintings. ``White noise'' in paintings would, for instance, consist of a canvas painted over with random shapes or splotches in random
colors at random locations on the canvas.


\section{Nature-beauty}

Suppose one is willing to accept the complexity criterion stated in the first law of aesthetic complexity.
Then one immediate question is about the quantitative amount of complexity required for an artistic design to be perceived as being ``beautiful.''
Stated differently, if art takes place in a ``bracket'' between monotony and irritation, between order and chaos, then what are the determining factors for the existence of this bracket and its extension?
Furthermore, can the bracket be enlarged? What determines its plasticity?

The \textit{second law of aesthetic complexity} states that:
\begin{quote}
{\em
Aesthetics are derived from natural forms.
}
\end{quote}

\begin{figure}
\centerline{\includegraphics[width=12cm]{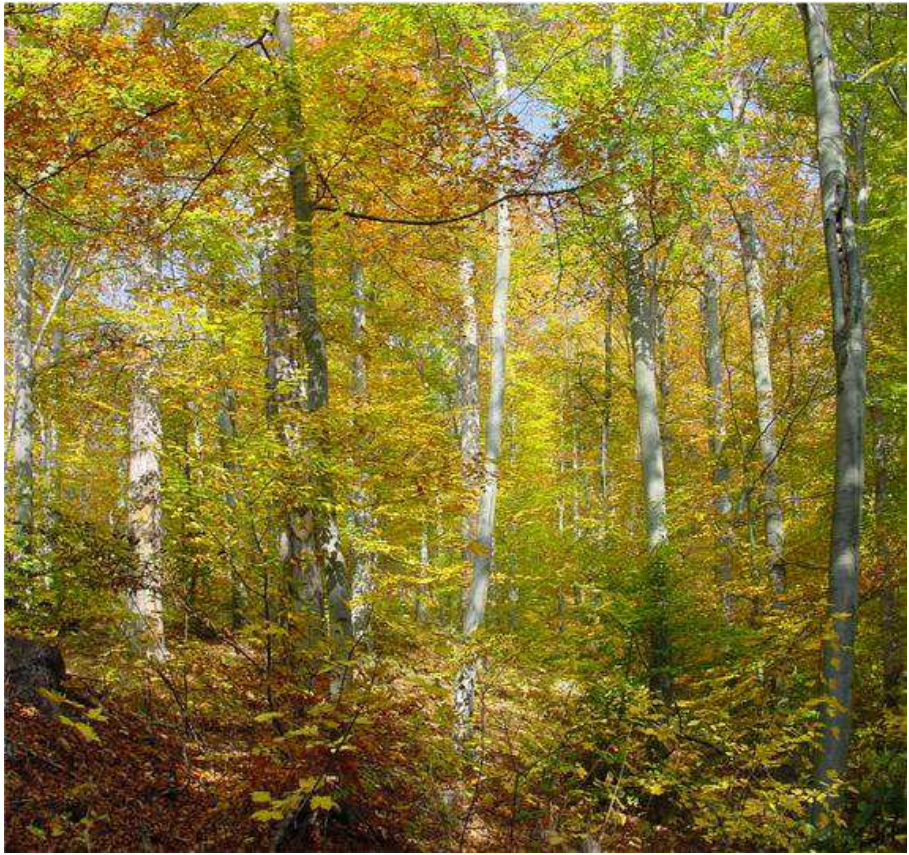}}
  \caption{Autumn foliage near Baden near Vienna, Lower Austria, Oct. 15, 2000
(\copyright Karl Svozil)}
   \label{2005-ae-foliage}
 \end{figure}

\begin{figure}
\centerline{\includegraphics[width=12cm]{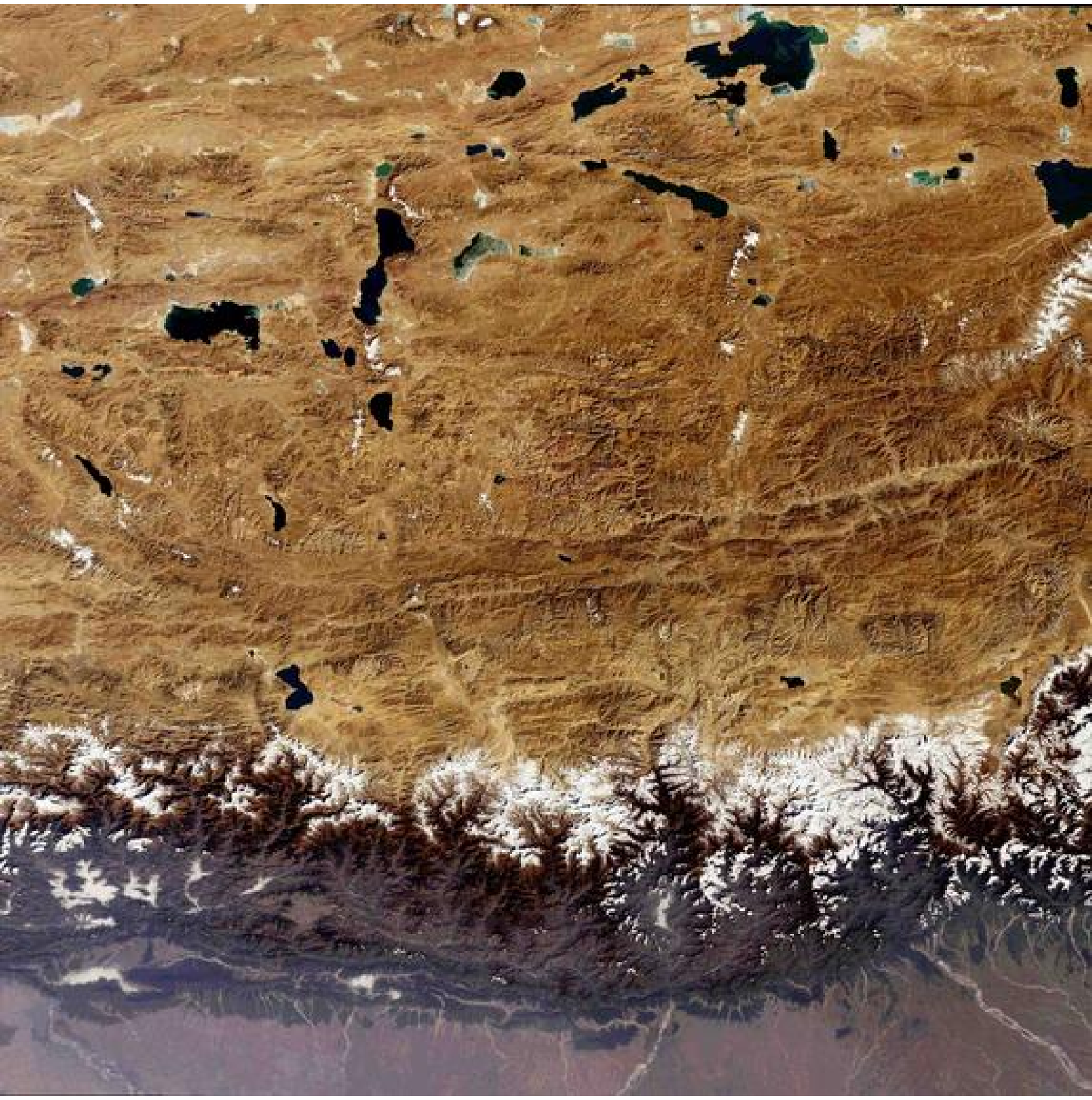}}
   \caption{Mount Everest as seen by MERIS at orbit \# 09148 on Nov. 30th, 2003
(\copyright ESA/MERIS)}
   \label{2005-ae-China_MountEverest_MER_FR_Orbit09148_20031130_hires}
 \end{figure}

The second law can be motivated both ontogenetically and phylogenically.
The human experience of art, at least where beauty and appreciative psychological responses are concerned, has been molded and shaped by the variations of natural forms such as clouds, rocks, leaves, waves, or the songs of birds.
Examples are depicted in Fig.~\ref{2005-ae-foliage}
\& \ref{2005-ae-China_MountEverest_MER_FR_Orbit09148_20031130_hires}.
Human artistic expression must cope with this human predisposition, which limits the plasticity and adaptability of human perception.
Regardless of the artistic idea, neglect of this condition may result in a sense of impertinence, provocation and ugliness for the person experiencing the creation.

Note that ``noise'' can be created relatively cheaply and simply, even though its algorithmic and computational complexity is high.
Very few lines of code are required to produce a statistically quasi-random scattering of shapes or color blotches.
For the artist willing to accept such automated techniques, this simplicity (``very few lines of code'') of protocol or technique means that the production of ``art'' is relatively effortless and cheap.
Typical examples from painting are ``splatter paintings'' (in German ``Sch\"uttbilder'') of some actionistic artists,
as well as music visualizations of some media player software.
The price being paid for the ease of creation is the loss of control and deliberation in the design, resulting in a sense of arbitrariness and incomprehensible superficiality on the receptive end.

By contrast, nature-beauty imposes heavy algorithmic costs on the creators of
virtual realities and arts in general, requiring a structural richness that
exceeds the power of contemporary computers by orders of magnitude.
To illustrate, let us consider how Nature herself creates nature-beauty. In
terms of algorithmic information content, it takes about 4 million
nucleotides (the basic molecules forming the nucleic acids DNA and RNA), and
about 4 thousand genes, to describe the simple bacterium Escherichia coli
(E. coli). This is the genome of E. coli, which for present purposes we may
equate with lines of code and functional segments of code.
Humans have about 1,000 times more nucleotides
than E. coli (around 3 billion), and an estimated 40,000 to 60,000 genes.
Every cellular entity on earth can be assumed to lie within those bounds.
The phenotype---that is the bodily creature --generated from these codes is
quite beyond the capability of contemporary computers. Even ``mere'' protein
folding remains one of the most difficult computational challenges of our
time. This compares indirectly to the exorbitant computational resources
needed to simulate an entire city in detail within a virtual reality.

Ultimately, the above theses will have to be tested against
experience and neurophysiological modeling. They relate, in some respects,
to Chomsky's system of transformational grammar. One of the possible tests
would be to differentiate between the ontogenetic and the phylogenetic parts
of the thesis. Children who grow up in rural surroundings might, for
instance, show very similar aesthetic preferences when compared to urban
children, although their environmental experiences vary widely. The same
should be true for people from very different environmental, cultural,
social and ethnic backgrounds.

\section{Aesthetic trends}

The third law of aesthetic complexity deals with its dynamics and with the changes in time of complexities in the various forms of artistic expressions:
\begin{quote}
{\em
Aesthetic complexity trends are determined by the available resources, and thus also by cost and scarcity.
}
\end{quote}

Artistic forms which are cost intensive, such as architecture or virtual realities, tend to become less complex.
Other art forms, such as painting and music, which may be augmented by ``cheap'' quasi-random methods of rendering, tend to become more complex by these methods.

One of the most influential critiques against man-made ornamentation was formulated in
1908 by Alfred Loos in his pamphlet ``Ornament und Verbrechen'' (English translation ``Ornament and Crime'')~\cite{loos}:
Loos argues that ornamentation is expensive; and resources diverted to decoration are wasted with regard to the functional value of the decorated objects.
Those resources could for instance be much better invested for leisure or for an increase in productivity.
{\em Loos' principle} can be pointedly stated by the following
question:
\begin{quote}
{\em
Why build one pretty house with ornamentation when you can have
two ugly ones for the same price?
}
\end{quote}

Such thoughts blended in well with Frederick Winslow Taylor's ``The
Principles of Scientific Management''
\cite{taylor-1911}  written in 1911
in the USA, as well as the socio-economic fantasies of the Bolsheviks in the USSR.
While such principles improved productivity and had a substantial impact on the growth of the economic output, they also increased the monotony of work and the human environment in general.
The reason for this was because Loos' principle has rendered an ideological and theoretical framework for the justification of low complex structures.

A mere programmatic commitment to ornamentation does not solve the problem of its cost, though.
After all, Loos did not criticize ornamentation {\em per se,} but the extra cost associated with it, which is not met by any immediately recognizable functional value.

\begin{figure}
\centerline{\includegraphics[width=12cm]{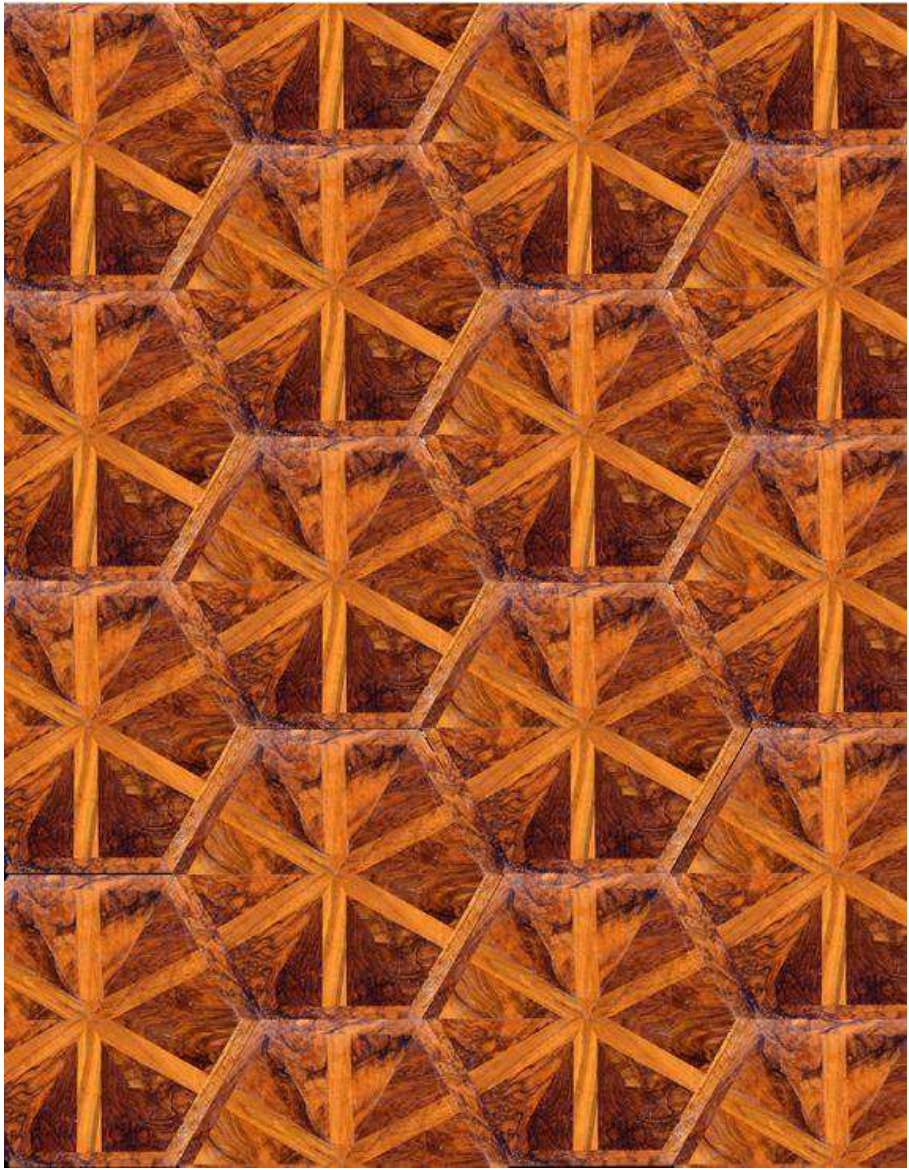}}
   \caption{Parquet flooring  in the galery rooms of the Garden Palais
 Liechtenstein, late 18th century, Vienna, Austria
(\copyright Sammlungen des F\"ursten von und zu Liechtenstein, Vaduz.
URL http://www.liechtensteinmuseum.at)}
   \label{2005-ae-flooring}
 \end{figure}

In acknowledging the need for ornamentation, Loos even suggested using naturally ornamented panels and templates such as wood or stone as a substitute for expensive human-crafted ornamentation.
Alas, natural ornamentation materials such as stones and wood are also expensive and not affordable by everyone (compare recent laminate floorings carrying photo reproductions of wood).
And as can be seen from the beautiful parquet flooring recovered recently in the Palais
Liechtenstein, Vienna, depicted in Fig.~\ref{2005-ae-flooring},
even laying natural panels
requires high craftsmanship and geometric sophistication.

The costs associated with aesthetics explain why only the rich and the aristocracy have had the privilege to live in abundantly decorated environments, with beautifully crafted ornaments and art throughout history.
Take the Roman villas,  the palaces of the renaissance and baroque periods as examples for an aesthetics affordable only to very few.
For the commoner, ornament and art has been hardly affordable throughout history.
One of the most efficient attempts to improve this situation was the production of bentwood furniture on a large scale by Thonet and Kohn industries around 1900.
Although since then, in some parts of the world the general living conditions have improved considerably, in this aesthetic respect, nothing has changed much:
the average citizen cannot afford beauty even today and lives in almost ridiculously styled environments mimicking ornamentation~\cite{koel-sack-dw}.

\section{Strategies to introduce richness at low cost}

Several strategies have been applied to increase the aesthetic complexity
and richness of artistic expression virtual universes. Many can also be found in nature. Some of
them are mentioned below. By automation, all these superficial strategies
may contribute towards the better acceptance of virtual realities and
ornamented forms in general without requiring too much human effort.
A word of warning seems not totally unjustified, though:
The human mind seems to be able to recognize automation in pattern creation, and often resents too simple schemes.

\begin{figure}
\begin{center}
\begin{tabular}{ccc}
 \includegraphics[width=5.23cm]{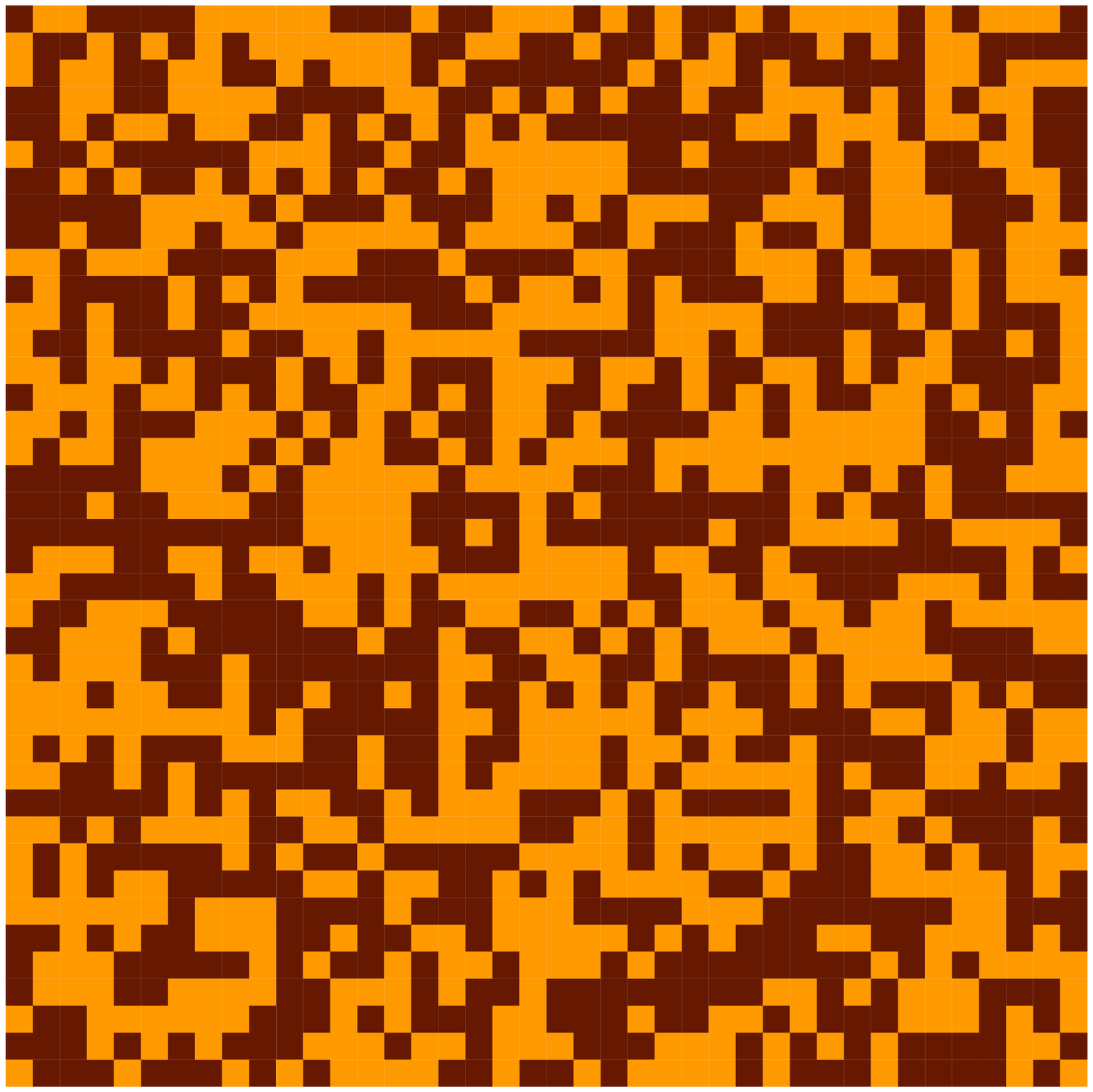}
&
 \includegraphics[width=5.23cm]{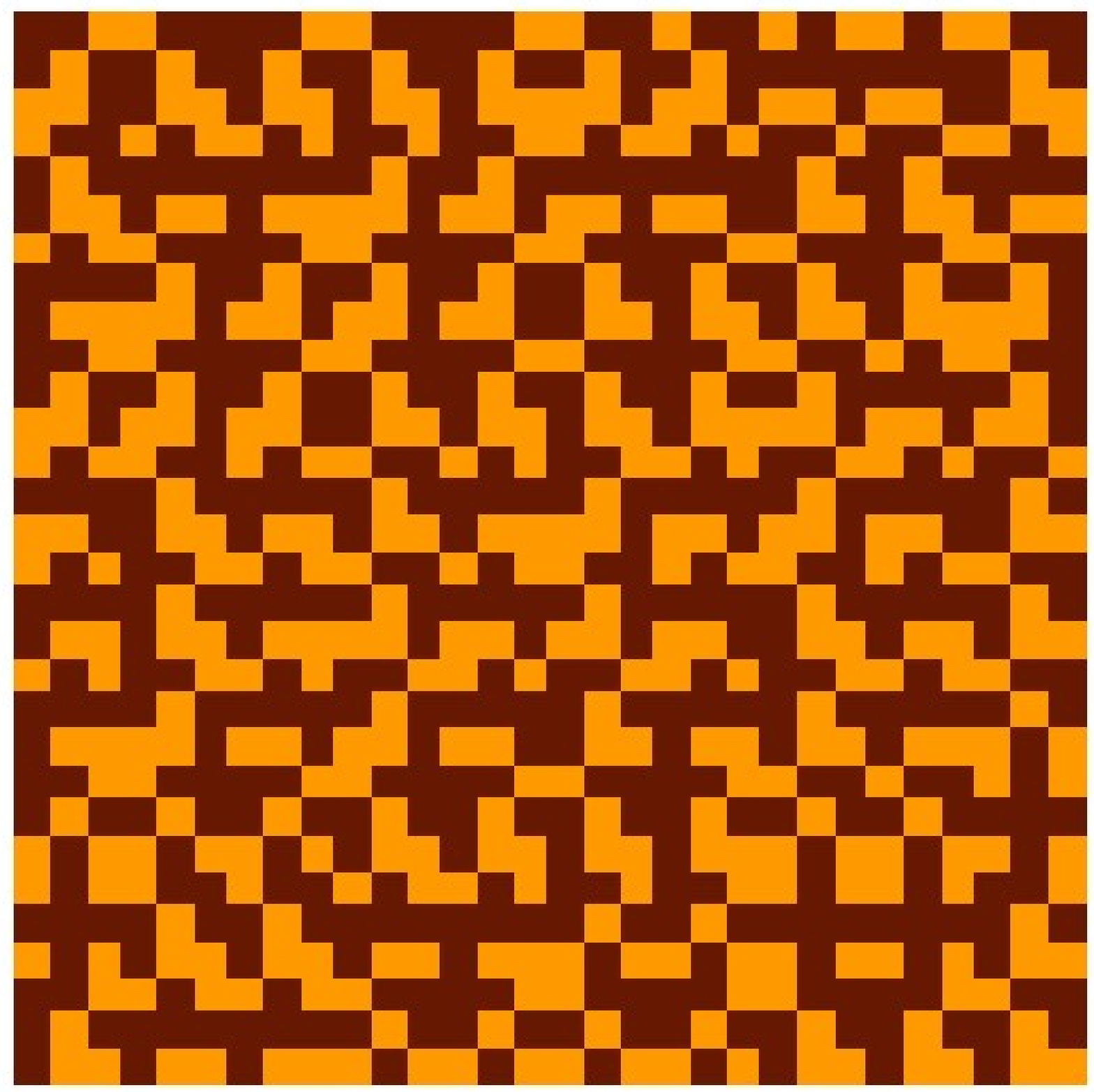}
&
 \includegraphics[width=5.23cm]{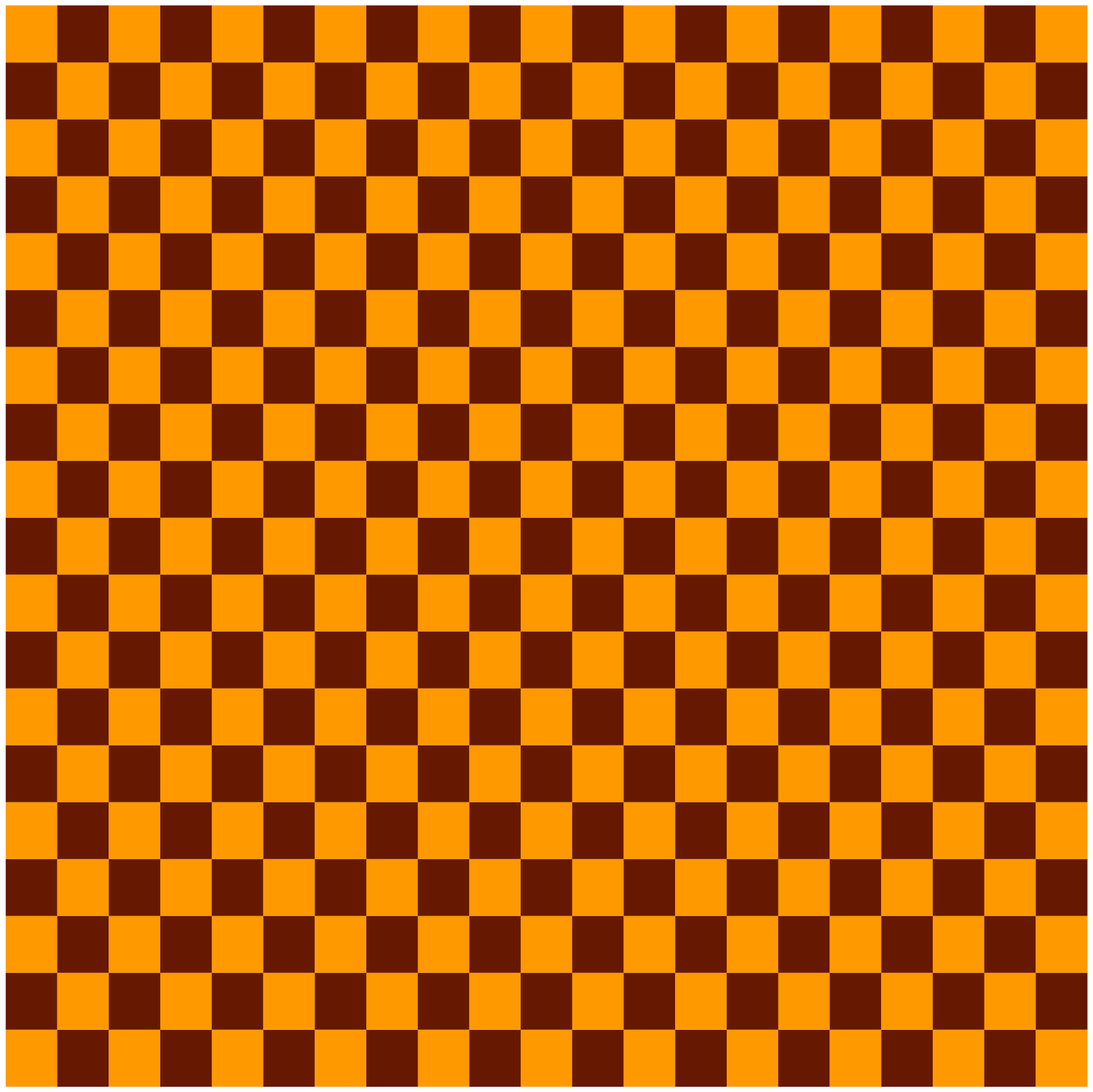}
\\
(a)&(b)&(c) \\
$\;$\\
 \includegraphics[width=5.00cm]{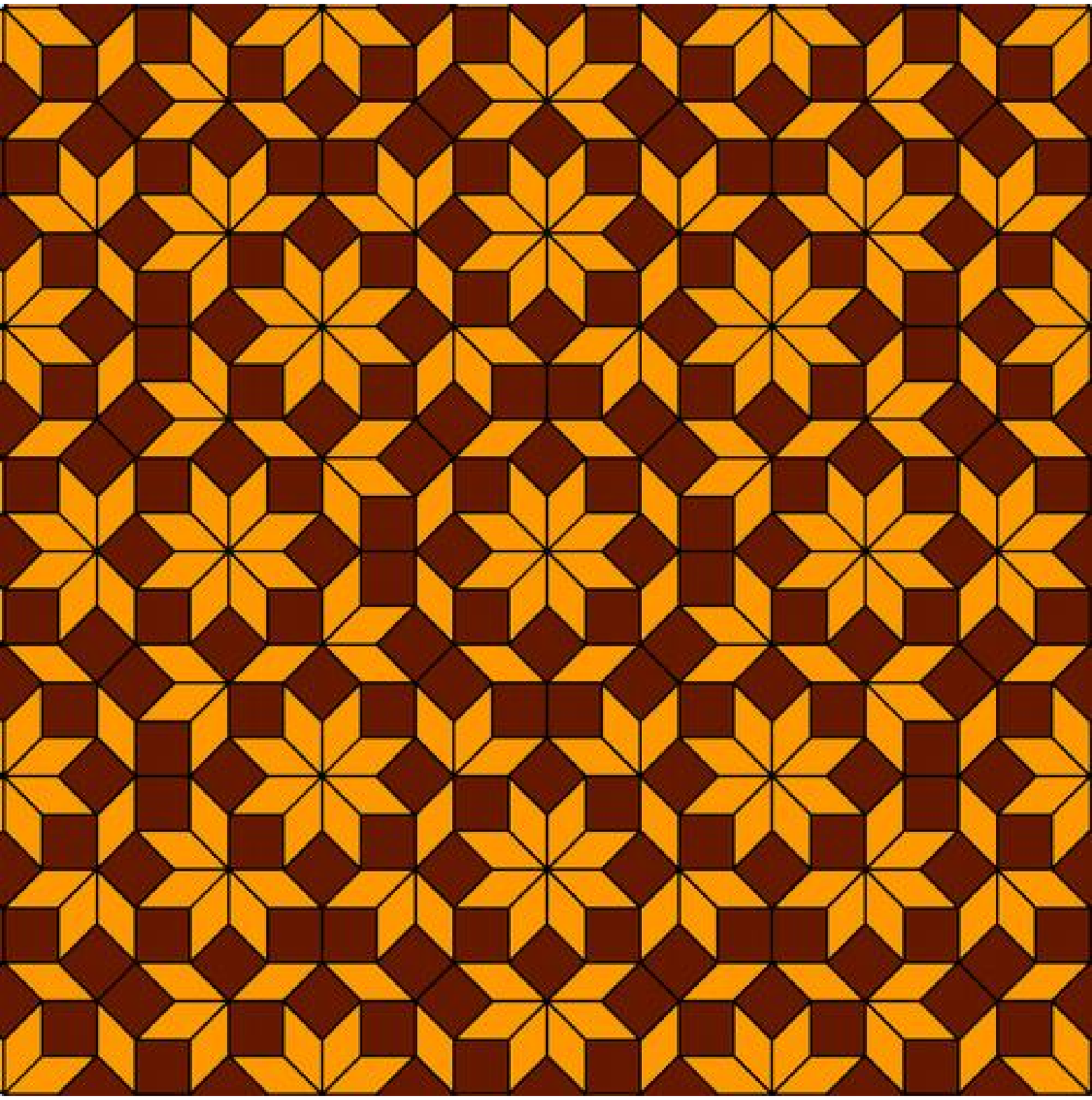}
&
 \includegraphics[width=5.00cm]{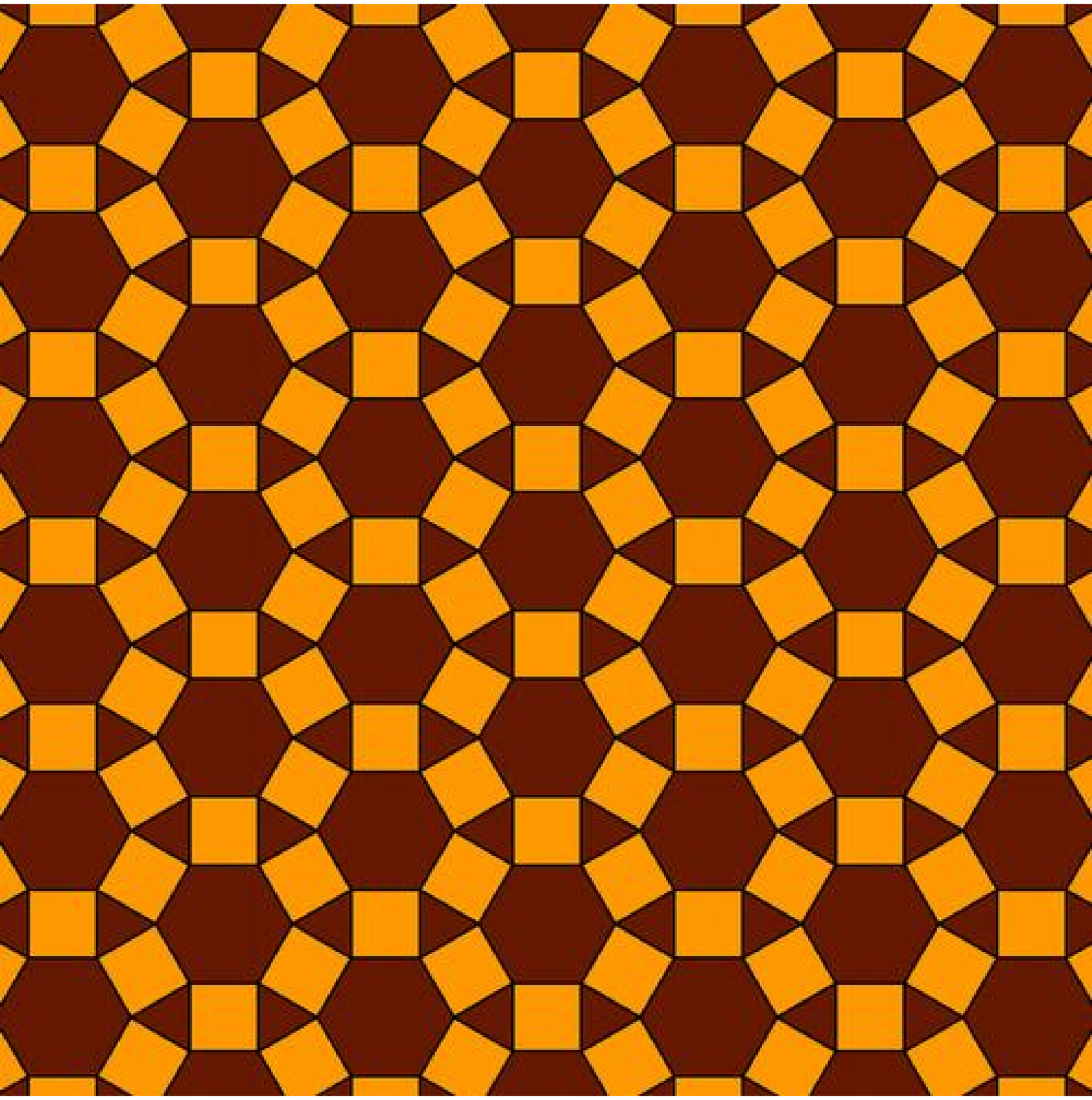}
&
 \includegraphics[width=5.00cm]{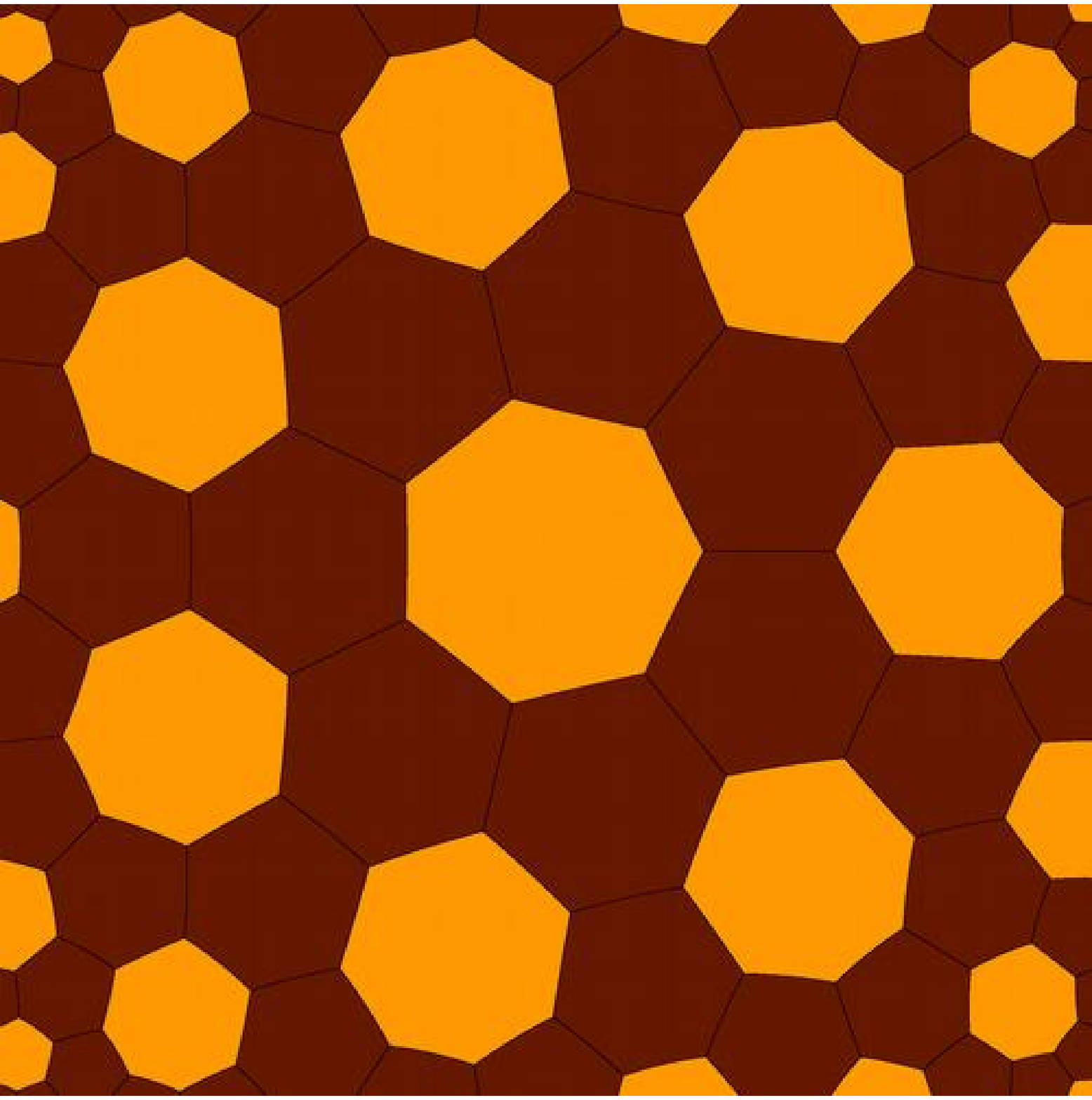}
\\
(d)&(e)&(f)\\
$\;$\\
 \includegraphics[width=5.00cm]{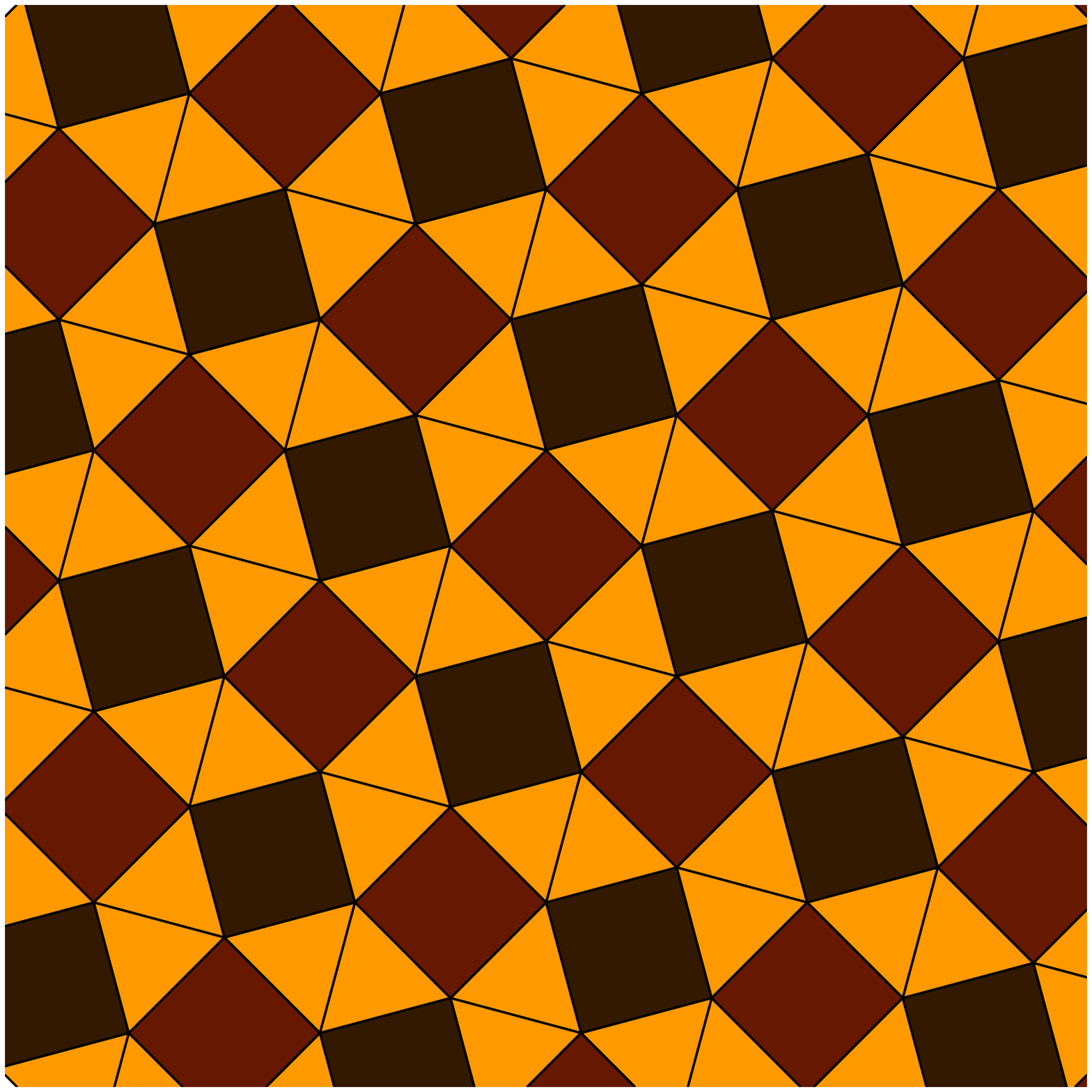}
&
 \includegraphics[width=5.00cm]{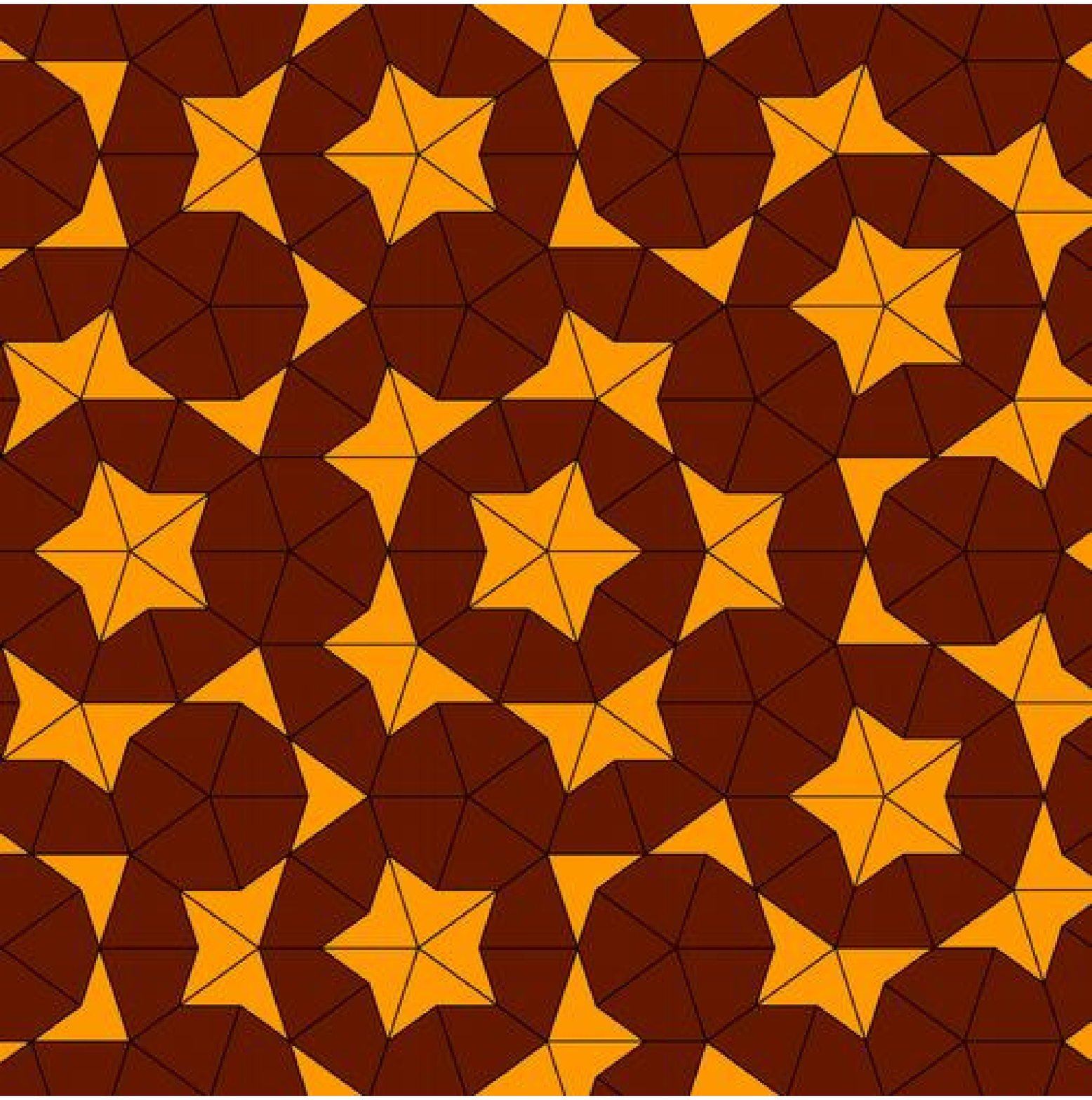}
&
 \includegraphics[width=5.00cm]{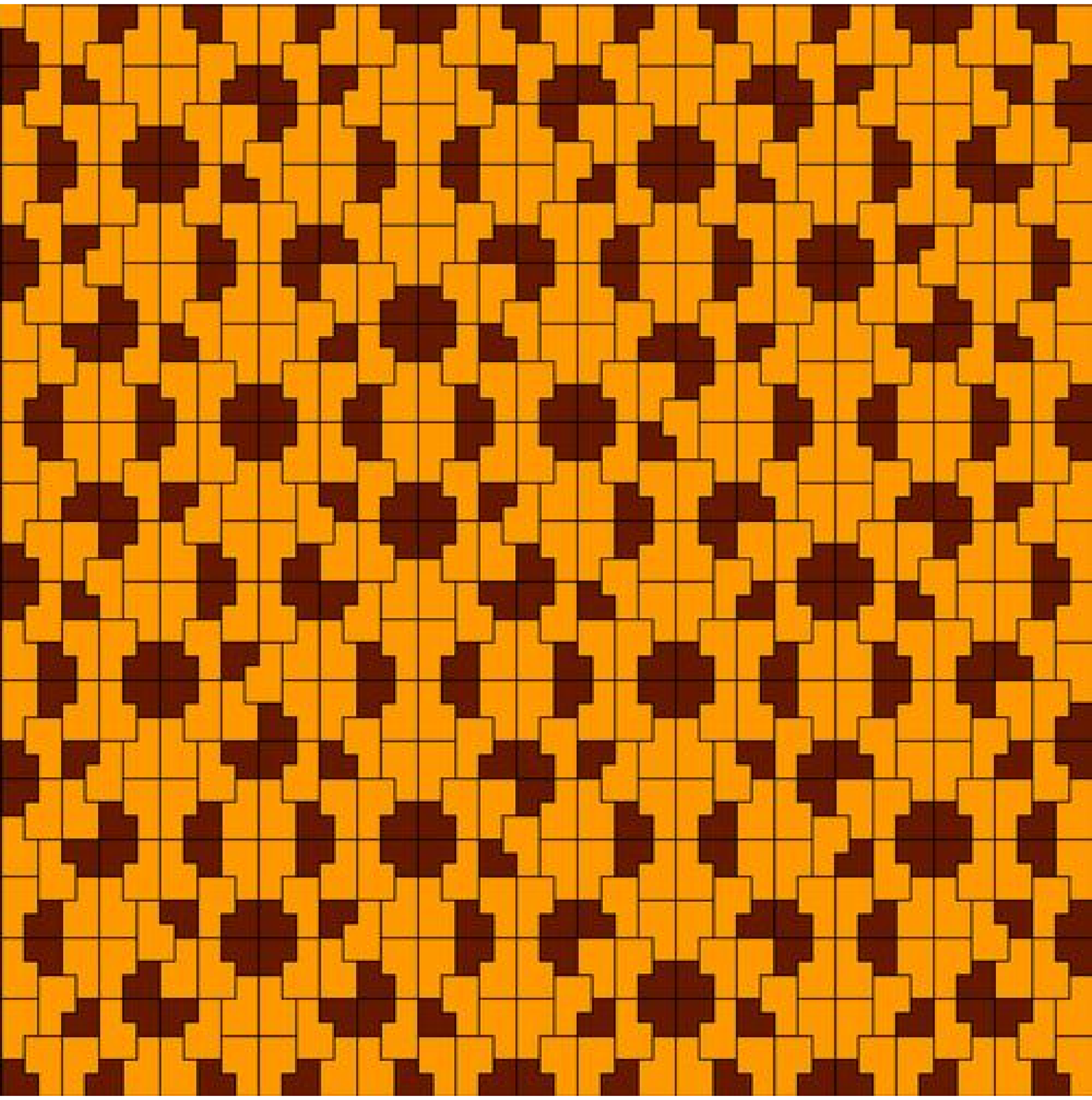}
\\
(g)&(h)&(i)\\
\end{tabular}
\end{center}
   \caption{Raster graphics
(a) from white noise;
(b) from permutations in a quantum~\cite{DonSvo01}
and automaton~\cite{svozil-2004-kyoto,svozil-2003-garda} state discrimination problem;
(c) from regular tessellation through repetition;
(d) Tiling obtained from the projection of a multi-dimensional hypercube with
an algorithm by  Grimm and Schreiber~\cite{grimm-schr-02};
(e-g) Tilings from
an algorithm by  Sremcevic and Sazdanovic (MathSource 4540);
(h) Tiling from
an algorithm by Lyman P. Hurd (MathSource 595);
(i) Ammann aperiodic tiling from
an algorithm by Sasho Kalajdzievski (MathSource 4273);
}
   \label{2005-ae-raster-wn}
 \end{figure}

\subsection{Randomness and mutation}

True randomness is a hypothetical postulated resource nobody knows to exist.
All ``algorithmic random number generators'' by definition produce
non-random output. Some random number modules have been proposed~\cite{svozil-qct}
and realised ~\cite{zeilinger:qct}
on the basis of physical processes such as quantum effects.
Yet, it can be safely asserted that for all practical purposes of
aesthetics, pseudo-random number generators suffice.

Alas, pure randomness is perceived as incomprehensible and irritating. For a
demonstration, the reader should contemplate the panel of random color
tiles in Fig.~\ref{2005-ae-raster-wn}(a).
Nevertheless, a certain randomization may improve the
perception of geometrical forms, making them appear ``less perfect'' and
``ideal'' by ``mutating'' them.

\subsection{Morphing and crossing of existing forms}

\begin{figure}[hptb]
\begin{center}
\begin{tabular}{c}
 \includegraphics[width=12cm]{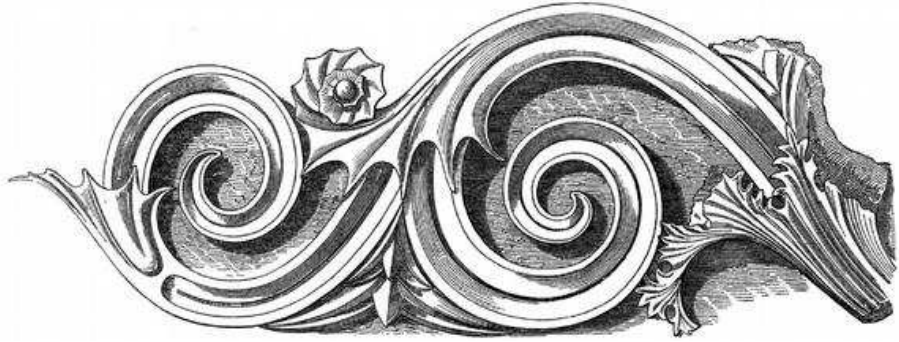}\\
(a)\\
$\;$\\
 \includegraphics[width=13cm]{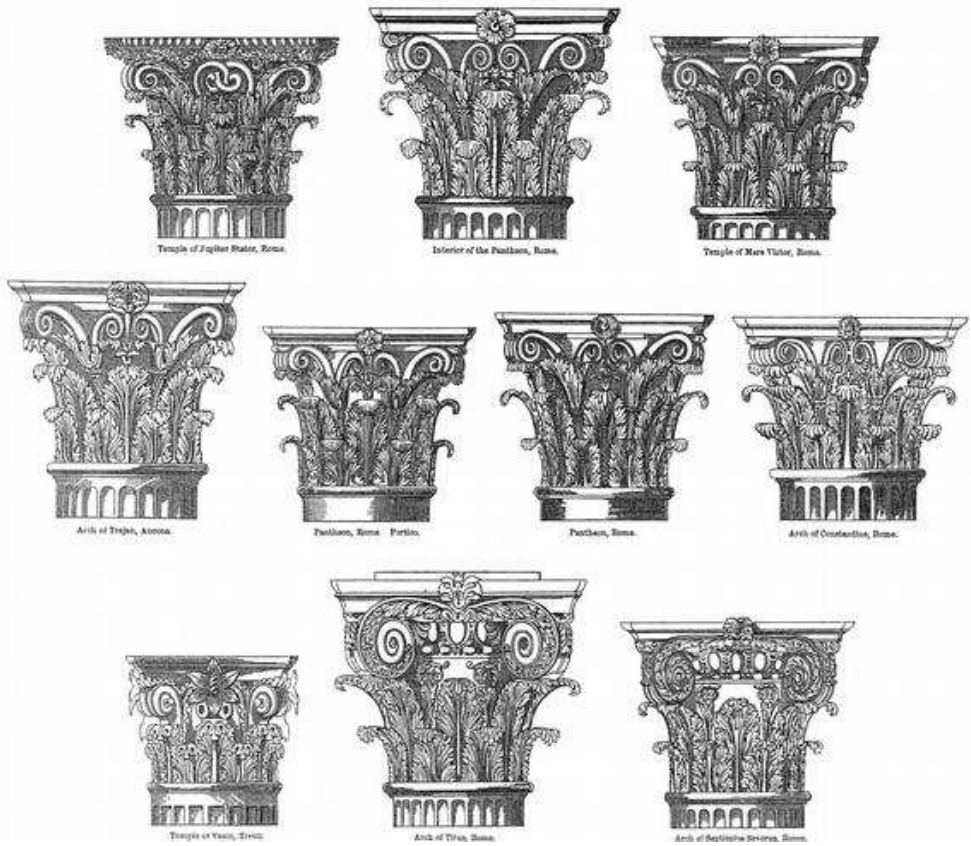}\\
(b)
\end{tabular}
\end{center}
   \caption{(a) Greek ornament from the Chorage Monument of Lysicrates, Athens; by Lewis Vulliamy and
reprinted by Owen Jones~\cite{jones-goo};
(b) Roman Corinthian and Composite Capitals reduced from Taylor and Cresy's {\it Rome}~\cite{Tay-Cr} and reprinted by Owen Jones~\cite{jones-goo};
}
   \label{2005-ae-owen}
 \end{figure}

This variation has been borrowed from Genetic Algorithms~\cite{goldberg:89,holland:92a,mitchell}.
It is
the deliberate use of natural forms such as leaves, trees, waves and so on,
morphing, crossing and blending them into existing functional and structural
entities.
Ancient civilizations such as the Greeks were masters of this technique.
The shape of Ionic and Corinthian Capitals,
as depicted in Fig.~\ref{2005-ae-owen}(b)
or the stucco detail in  Fig.~\ref{2005-ae-bospiral}
are such examples.
\subsection{Permutation}

Permutations are a means to repeat one and the same formal message over and
over again without repeating it syntactically. Strictly speaking, it should
be considered in the symmetry section below. One of the decisive features of
permutations are the reversibility, the ``one-to-one-ness'' of the
associated transformations.
Fig.~\ref{2005-ae-raster-wn}(b) depicts a permutation pattern previously generated
in the context of quantum state discrimination
\cite{DonSvo01,svozil-2002-statepart-prl,svozil-2004-kyoto,svozil-2003-garda}.

\subsection{Self-similarity}

Self-similar ``fractal'' \cite{mandelbrot-77,mandelbrot-83,falconer1,falconer2}
geometries have been discussed intensively in conjunction with authentically looking landscapes \cite{voss85} and architectural forms~\cite{bovill,jencks},
as well as music~\cite{gard-78} and paintings~\cite{taylor-99}.
As demonstrated by the image compression techniques from iterated functions systems~\cite{barnsley:88},
fractals are generated by the successive iteration of certain non-linear mappings.

It should be realised however, that although fractal forms abound in nature, their virtually generated doubles often tend to appear boring and artificial.
A combination of fractal symmetry and random mutation may be a good recipe for creating interesting patterns.

\subsection{Repetition}

Repetition of patterns and reproduction of natural forms such as the ones in Fig.~\ref{2005-ae-raster-wn}(c,e,g) may be a great design resource.
It should be noted that without any modifications such as mutation, the repetition of small structures can be decoded very easily and thus may appear monotonous.
One should, however, not underestimate the joy people experience by listening to something they already know \cite{feynman-law}!

\subsection{Symmetry}

Ornamentation by symmetric patterns is an ancient method.
Contemporary mathematics offers a pandemonium of different symmetric patterns~\cite{gruenbaum-tiling},
the formally most advanced being aperiodic tilings~\cite{baake-02,grimm-schr-02}.
Figs.~\ref{2005-ae-raster-wn}(d,f,h,i) depict such aperiodic floor tilings.
These tilings would not have been possible a few years ago and therefore are not realized in any historic building.




\begin{figure}
\centerline{\includegraphics[width=12cm]{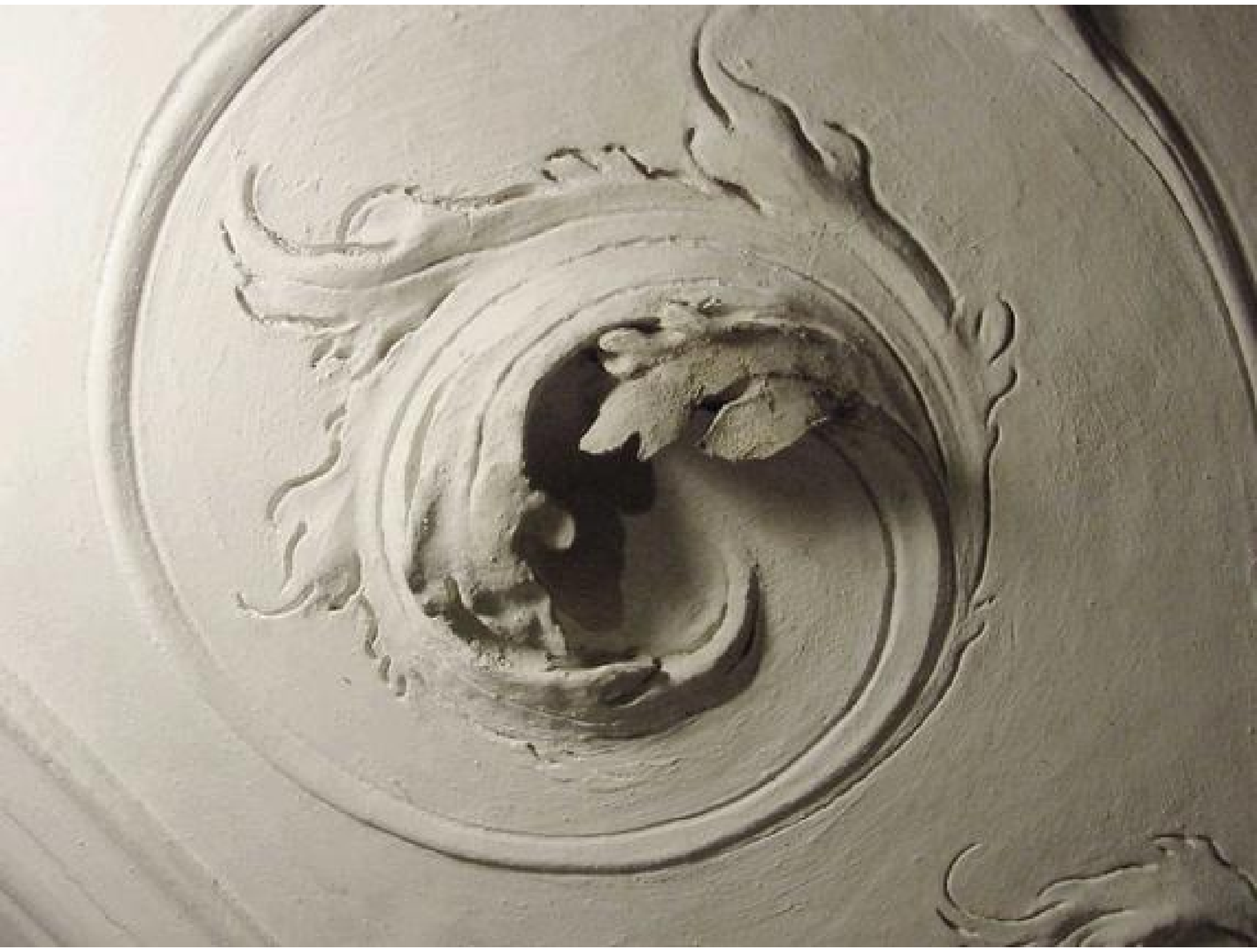}}
   \caption{Santino Bussi (1664-1736) Stucco detail in the Sala Terrena of the Garden Palais
 Liechtenstein, after 1700, Vienna, Austria
(\copyright Sammlungen des F\"ursten von und zu Liechtenstein, Vaduz.
URL http://www.liechtensteinmuseum.at)}
   \label{2005-ae-bospiral}
 \end{figure}

\section{Summary and outlook}

Aesthetics has been examined from the point of view of statistical complexity measures of actions, performances and renditions required for creating and decrypting a work of art.
Whereas a too condensed ``highly complex'' encoding makes a decryption of a work of art impossible and is perceived as chaotic by the untrained mind,
an encoding of ``low complexity'' results in too regular structures which are perceived as monotonous, too orderly and not very stimulating.
There seem to exist bounds from above and from below on artistic expression:
art can neither exist in a scheme dominated by chaos, randomness, arbitrariness and white noise, nor can it exist in a regime dominated by too much order, monotony and dullness.
Hence, a necessary condition for an artistic form or design to appear appealing is its complexity to lie within a bracket between monotony and chaos.
It has been argued that, due to human predisposition, this bracket is invariably based on natural forms; with rather limited plasticity.
We have also observed that historically aesthetic complexity trends are dominated by cost and scarcity.
Thus painting and music tend to become ``enriched'' with cheap random elements, causing these artistic forms to appear incomprehensive and meaningless;
whereas architectural forms tend to become less complex, rendering an overal feeling of boredom, dullness and dreariness.

To overcome these issues, we have argued for the necessity of ornamentation, decoration and the presence of nature-beauty as a precondition for aesthetic acceptance.
Thereby, we have in mind statistical and algorithmic complexity measures and methods to evaluate and automatically generate ornamental forms and designs.

Nothing has been said about human originality and artistic talent.
Indeed, the more one attempts to argue for the necessity and feasibility of
automated creation of ornamentation in accord with nature-beauty, the more
it becomes clear how brilliant, gratifying and truly enjoyable human
artistic expressions can be.

\begin{figure}
\centerline{\includegraphics[width=12cm]{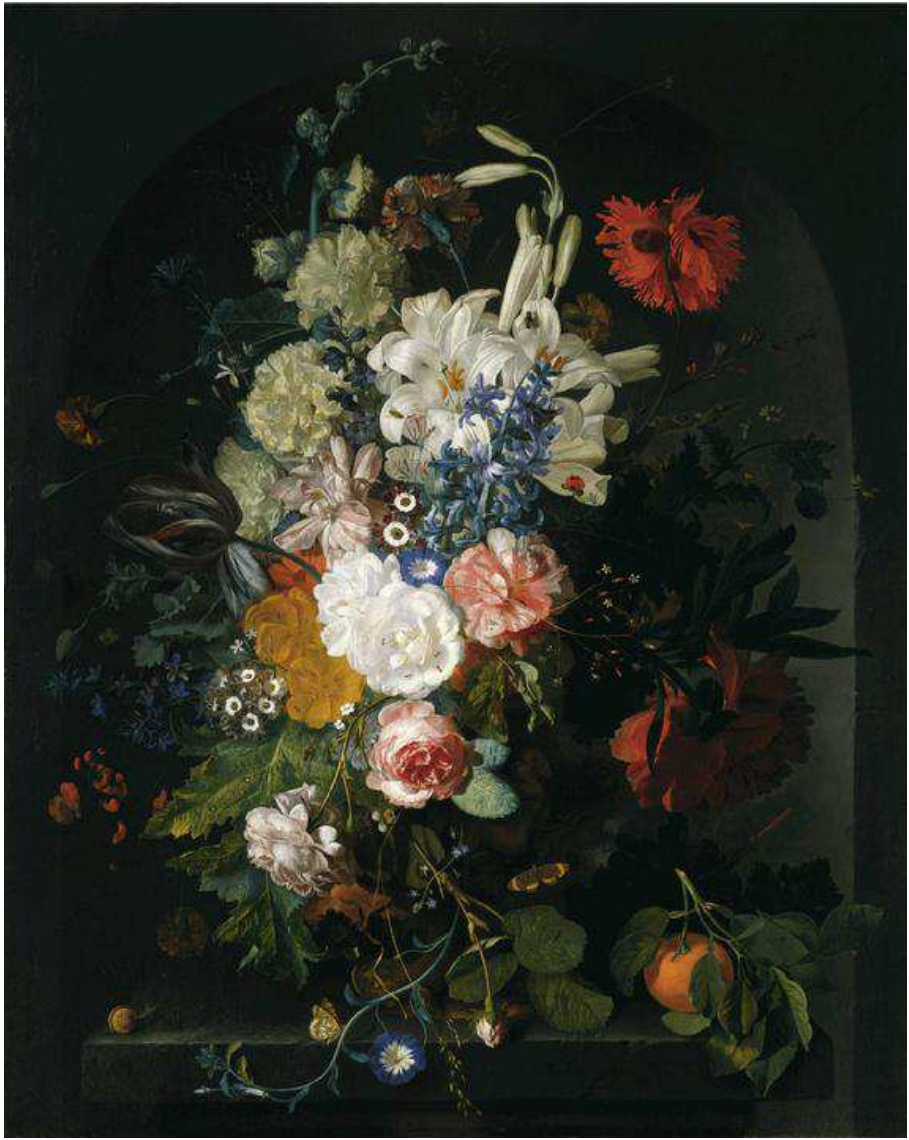}}
   \caption{Jan Van Huysum, Flowers
(\copyright Sammlungen des F\"ursten von und zu Liechtenstein, Vaduz.
URL http://www.liechtensteinmuseum.at)}
   \label{2005-ae-JanVanHuysum_Blumenstrauss}
 \end{figure}

Consider, for example, the traditional ornaments collected by Owen Jones
\cite{jones-goo} and depicted in Fig.~\ref{2005-ae-owen}, the stucco created by Santino Bussi and
depicted in Fig.~\ref{2005-ae-bospiral}, and Jan Van Huysum's bouquet of flowers
in Fig.~\ref{2005-ae-JanVanHuysum_Blumenstrauss}.
Very often, books of fauna, mushrooms and botany in general rely on human drawings rather than on photography in order to be able to properly depict species.
It may appear even questionable whether the automation of
pattern formation will ever be capable to fully substitute or outperform
human art.
One is reminded of similar debates in artificial intelligence research.

\section*{Acknowledments}
I would like to express my special thanks to Ross Rhodes
for critically reading and revising the manuscript.
Thanks go also to (in lexicographic order)
Tim Bo,
Cristian Calude,
Georg Franck-Oberaspach,
G\"unter Krenn,
Volkmar Putz,
Christian Schreibm\"uller,
and
Udo Wid
for discussions and references.
Many ideas grew from a research cooperation with Klaus Ehrenberger
of the Medical University of Vienna on the coding and processing of
stimuli by cochlear implants, a direct artificial interface to the cortex.
Almost needless to say,
I take full responsibility for controversial statements.
Reproductions of Figs.~\ref{2005-ae-flooring},~\ref{2005-ae-bospiral}, and~\ref{2005-ae-JanVanHuysum_Blumenstrauss}
with kind permission of the Sammlungen des F\"ursten von und zu Liechtenstein.


\end{document}